\documentclass[runningheads,a4paper]{llncs}
\authorrunning{Philipp Jordan et al.}
\usepackage[utf8]{inputenc}
\usepackage{amssymb}
\usepackage{graphicx}
\usepackage[normalem]{ulem}
\useunder{\uline}{\ul}{}
\usepackage{subfig}
\usepackage[strings]{underscore}
\usepackage[hidelinks]{hyperref}
\hypersetup{
    colorlinks=true,
    linkcolor=blue,
    filecolor=magenta,      
    urlcolor=cyan,
}
\title{Exploring the Referral and Usage of Science-Fiction in HCI Literature\thanks{The final publication is
available at \url{https://link.springer.com/chapter/10.1007\%2F978-3-319-91803-7_2}}}
\author{Philipp Jordan\textsuperscript{1} \and Omar Mubin\textsuperscript{2} \and Mohammad Obaid\textsuperscript{3} \and Paula Alexandra Silva\textsuperscript{4}}%
\institute{\textsuperscript{1} University of Hawaii at Manoa, Honolulu, USA\\
\email{philippj@hawaii.edu}\\
\textsuperscript{2} SCEM, Western Sydney University, Sydney, Australia\\
\email{o.mubin@westernsydney.edu.au}\\
\textsuperscript{3} Uppsala University, Uppsala, Sweden\\
\email{mohammad.obaid@it.uu.se}\\
\textsuperscript{4} DigiMedia Research Center, University of Aveiro, Aveiro, Portugal\\
\email{palexa@gmail.com}}
\begin{document}
\maketitle
\begin{abstract}
Research on science fiction (sci-fi) in scientific publications has indicated the usage of sci-fi stories, movies or shows to inspire novel Human-Computer Interaction (HCI) research. Yet no studies have analysed sci-fi in a top-ranked computer science conference at present. For that reason, we examine the CHI main track for the presence and nature of sci-fi referrals in relationship to HCI research. We search for six sci-fi terms in a dataset of 5812 CHI main proceedings and code the context of 175 sci-fi referrals in 83 papers indexed in the CHI main track. In our results, we categorize these papers into five contemporary HCI research themes wherein sci-fi and HCI interconnect: 1) Theoretical Design Research; 2) New Interactions; 3) Human-Body Modification or Extension; 4) Human-Robot Interaction and Artificial Intelligence; and 5) Visions of Computing and HCI. In conclusion, we discuss results and implications located in the promising arena of sci-fi and HCI research. 
\keywords{Design fiction, Future visions, HCI inspiraton, Popular culture in science, Science fiction}
\end{abstract}

\section{Introduction}
Sci-fi has inspired many technological innovations. Its influence is present in the invention, conceptualization, design and application of interfaces and technology. As an avenue of creativity and expression both, sci-fi literature and media have fueled advancements in interactive technology and proven to be a key source of inspiration for researchers in the field of computing technology \cite{asimov1975easy}.

There are numerous examples of interactive products, devices and systems in the real world whose origin can be traced back to sci-fi \cite{Jordan2018,Mubin:2016}. From the wristwatch used by fictional detective Dick Tracy to the communicators of \textit{Star Trek} which predated nowadays mobile phones. From the video conferencing and disobedient Artificial Intelligence (AI) as depicted in \textit{2001: A Space Odyssey}, to the video-phones and robots of Fritz Lang's 1927 \textit{Metropolis}, both considered influential and seminal sci-fi dystopian movies to this day. 
Upon closer investigation, we observe that this relationship is bi-directional to such an extent that sci-fi creators regularly consult with researchers and scientists \cite{kirby2009future}, for instance Marvin Minsky's \cite{marvin2016} contribution to \textit{2001: A Space Odyssey}. The areas of HCI and sci-fi therefore seem to have the potential to grow in unison \cite{marcus2013}. 

In summary, it is acknowledged that sci-fi has had an impact on HCI, computing and interactive technology. This is confirmed in a number of overview and survey articles. Aaron Marcus has led and organized a number of CHI panels on the topic, in particular in 1999 \cite{marcus1999opening} where he listed a set of key reflective questions intended to dissect the relationship between HCI and sci-fi. Unfortunately, this and other similar works only provide a meta-level understanding of this relationship by stating an impact of some extent, such as case studies on collaborations of researchers and movie-makers  \cite{kirby2009future} or loose anecdotal evidence. Hence, there seems to be a lack of descriptive data and qualitative assessment of how the relationship between sci-fi and HCI has evolved over the years in HCI research.
 
But how does the anecdotal, discursive symbiosis of HCI and sci-fi translate into scientific research and peer-reviewed publications? Our research question examines how sci-fi is used in a leading HCI venue, in this study, the ACM Conference on Human Factors in Computing Systems (CHI). Specifically, we explore what particular kinds of sci-fi HCI researchers mention in their papers and how these sci-fi referrals relate to the research itself. At this stage, we limit our investigation to the scientific publications produced in the 
main proceedings of CHI. Given the high ranking of this particular conference \cite{confportal,GSconfportal}, and impact of the main proceedings (e.g. 10 of 10 of the highest cited papers of the CHI conference are indexed in the main CHI proceedings \cite{ACMDL}), we exclude the CHI extended abstracts in this preliminary study.


In the remaining sections of the paper, we first present a summary of related work on the topic followed by an account of our process of sci-fi keyword analysis through a contextual analysis of HCI research. We then present our coding scheme, overviews, and details of our results and speculate on their meaning. In our future work, we outline the next steps of this research.

\section{Background}
Contemporary HCI research is gradually beginning to disseminate the past, present, and future value and impact of sci-fi visualizations on HCI research and vice-versa \cite{Jordan2018,Jordan2016,Marcus:2015,Mubin:2016}. Among others, various case studies on the intersections of HCI and sci-fi movies \cite{Larson2008,Schmitz:2008}, on novel sci-fi movie-based interaction techniques \cite{Figueiredo:2015:OCH:2702613.2732888}, sci-fi inspired future user interfaces and design heuristics \cite{Shedroff:2012,Troiano:2016} have been published. In a broader context, sci-fi prototyping \cite{Kohno:2011} 
and the lately in vogue emergence of design fiction as a method in design research (e.g. \cite{Lindley:2015}) substantiate a mutual relationship of HCI and sci-fi. 

While it is acknowledged that HCI and interaction design can learn from sci-fi \cite{Marcus:1992,marcus1999opening}, the HCI community is realising that there is more to sci-fi's relationship with interactive technology than presenting an utopian vision; which is unfortunately what most overview papers get drawn into. Most HCI/sci-fi reviews \cite{Schmitz:2008} simply list sci-fi technologies that have been realized in the real world; the HCI community might find it difficult to utilize such insights to advance their research. Supplementing such overviews through sci-fi related search-and-retrieval queries in computer science conferences can facilitate a better description of the intersection and utilization of HCI and sci-fi; and therefore explain how HCI researchers actually use sci-fi in their research publications.

To create a better and more nuanced understanding of the sci-fi and HCI relationship in the case of scientific publications, this paper presents a preliminary study, where we collect and qualitatively analyse the  presence of sci-fi in the CHI main proceedings, one of the most prestigious, contemporary HCI collections. By doing this analysis, we expect to gain a better understanding of the evolutionary, historic, and chronological pattern in the presence of sci-fi related referrals in HCI research.As prior research \cite{Liu:2014} has shown that CHI proceedings are prone to general topic shifts over extended periods of time, we do expect to uncover a similar shift towards sci-fi related referrals in the most recent CHI publications due to the variation in the popularity and propagation of particular sci-fi content.

An important parameter to judge the influence of any discipline on a field of research is the analysis of the research articles emerging from that area. Scientometrics, keyword- and citation-analyses \cite{MINGERS20151} are defined as the area of research focusing on measuring and analyzing the impact of various factors on science and technology. The technique is now widely utilized in understanding publication trends and emerging fields in HCI. For example, we identify a variety of scientometric studies on HCI literature \cite{henry2007}, and specifically the CHI conference \cite{Bartneck:2009:SAC:1518701.1518810,Guha:2013:FBF:2467696.2467732,Liu:2014}. Nevertheless, aside of one example \cite{levin2014films}, scientometric or topical studies that analyze the presence of sci-fi related terms and keywords in HCI or CHI literature seem to lack. 

Hence, we believe that an analysis of the presence and referrals of sci-fi related keywords in the CHI main track is an important first step to determine a premature, but much more informed and data-based insight into the relationship of the two disciplines of HCI and sci-fi.The earlier introduced exception is Levin's \cite{levin2014films} study on the presence of 20 sci-fi movies in the ISI web of science database. While this study is methodologically the closest related state-of-the-art to our study, it differs in two  significant aspects: 

Firstly, Levin \cite{levin2014films} assessed the overall presence of these sci-fi movies across a broad range of research disciplines, for instance aerospace engineering, economics, political sciences and biology thereby extending beyond our current scope and research aim. Consequently, as Levin's study extended into such a wide variety of research topics, computing technology was under-represented. 

Secondly, Levin's study \cite{levin2014films} as well as our own research \cite{Jordan2018} on the usage usage of \textit{Star Trek} in the ACM Digital Library, do both focus on very specific and selected sci-fi movies and franchises, e.g. \textit{Blade Runner}, \textit{Matrix} in the former and \textit{Star Trek} in the latter study. We therefore assess in this study the general presence and usage of unspecific ``sci-fi'' in HCI publications through a more generic, inclusive search-query. 

In the remaining part of the paper, we do summarize in section 3 our method and data collection process, describe in  section 4 the results of our study,  summarize  the findings and conclusions in section 5 and outline the future work in section 6.






\section{Method}

\begin{figure}[ht]
  \centering
  \includegraphics[scale=0.33]{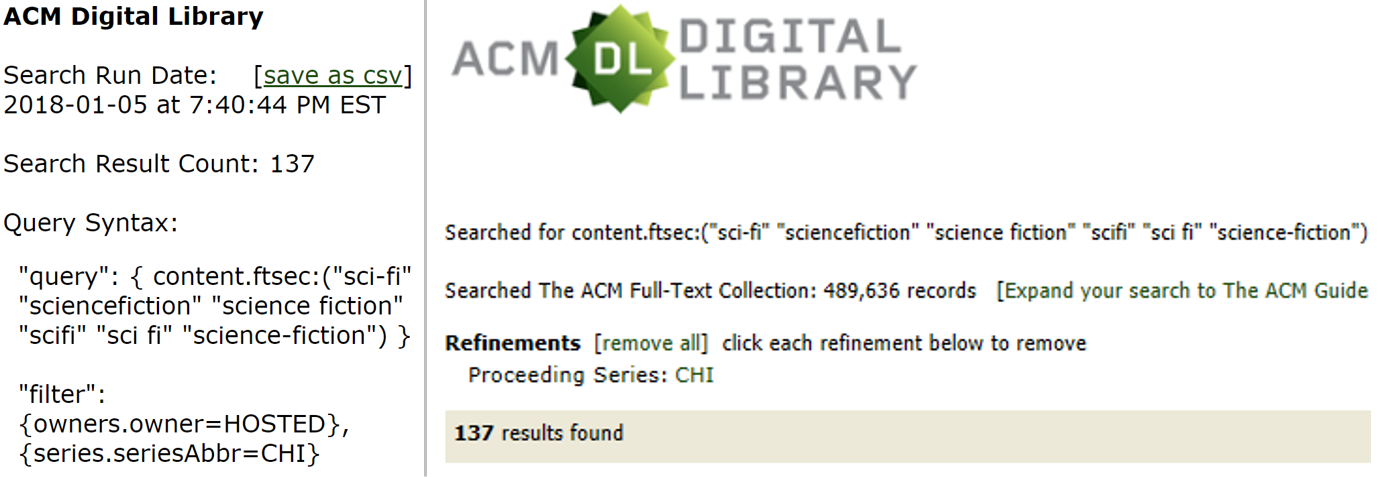}
  \caption{ACM Digital Library search query and results}
  \label{fig:acmsearch}
\end{figure}

Using the ACM Digital Library search interface, we query the full-text of the entire CHI proceedings in early 2018 for each of the six search terms: \textit{``sci-fi'' ``sciencefiction'' ``science fiction" ``scifi'' ``sci fi'' or  ``science-fiction''}, see also Figure \ref{fig:acmsearch}. 
\subsubsection{Search query}
\begin{verbatim}
Search Run Date: 2018-01-05 at 7:40:44 PM EST
Search Result Count: 137
Query Syntax:
"query": { content.ftsec:("sci-fi" "sciencefiction" 
"science fiction" "scifi" "sci fi" "science-fiction") }
"filter": {owners.owner=HOSTED},
{series.seriesAbbr.CHI}
\end{verbatim}
This query searches  all records in the CHI conference in the ACM Digital Library and returns 137 records in January 2018. These 137 retrieved records can be categorized into 83 records indexed in the main CHI proceedings (CHI PRO) and 54 indexed in other CHI series (CHI OTH), see also Table \ref{tab:acmsearch2}. Among those papers in the CHI OTH series, we find the CHI extended abstracts, work-in-progress papers or panel-, interactivity- or student-sessions. In this preliminary study, we qualitatively analyze the 83 records retrieved in the main CHI proceedings.



First, we conducted an open coding process of the 83 records in the CHI main proceedings amid the authors. This open coding is based on the specifity of the sci-fi referral and the sci-fi referral relationship to the primary technologies, respectively the main research theme of the individual publication. To both, establish initial coding categories and ensure a basic reliability of these categories, each author first independently reviewed 15 papers of the 83 papers. Second, a group review of these 15 papers and codes advanced the coding scheme outlined below. Third, all authors then proceeded to code the remaining 68 papers using the coding scheme and, fourth, through a final group review did resolve any remaining inconsistencies. In detail, our coding scheme consists of three main items:

\begin{enumerate}
\item \textbf{Unspecific sci-fi referral} - describes the unspecific sci-fi refereed in the retrieved publication, such as a general referrals to sci-fi movies, television shows, sci-fi short stories, books or literature in the full-text, for example \textit{``science-fiction stories''} or \textit{``sci-fi movies''}.
\item \textbf{Specific sci-fi referral} - if applicable, describes the details of the sci-fi media type referred in the retrieved publication, such as the specific title of a sci-fi movie, show, book or an explicitly mentioned name of a sci-fi movie or book author, for example \textit{``2001: A Space Odyssey''} or \textit{``Bruce Sterling''}.

\item \textbf{Sci-fi / HCI relationship} - if applicable, describes  the primary technologies or concepts from a sci-fi movie, show, or story in relationship to the research theme and focus of the individual, retrieved publication. Examples are herein \textit{``shape-changing interfaces''}, \textit{``autonomous cars''} but also theoretical papers on \textit{``futurism''} and \textit{``design research''}.
\end{enumerate}




\section{Results}
\begin{figure}[h]
  \centering
  \includegraphics[scale=0.45]{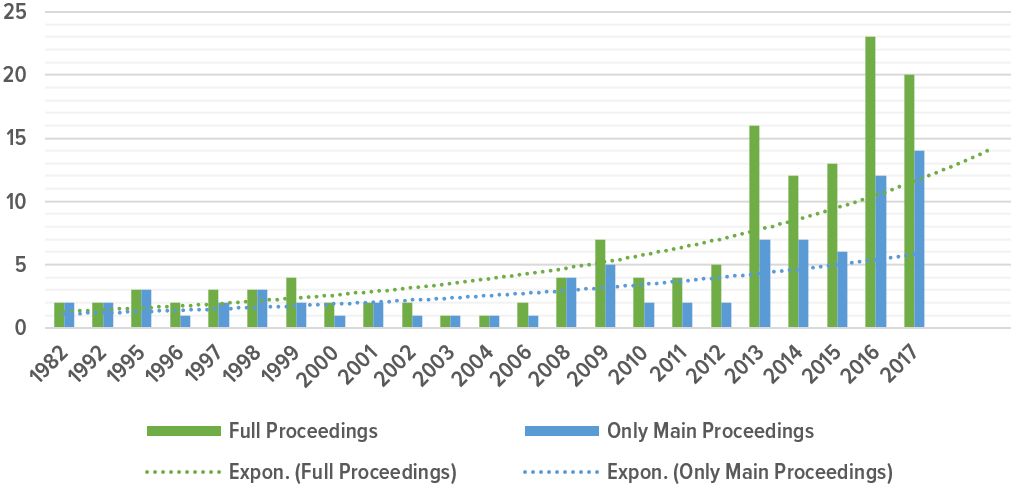}
  \caption{Publications matching our sci-fi query over the full CHI proceedings (green bars, n=137), and the main CHI proceedings (blue bars, n=83).}
  \label{fig:results1}
\end{figure}

Our query, as defined in Figure \ref{fig:acmsearch}, retrieves 137 records distributed into 83 records in the CHI main proceedings and 54 records in the other CHI proceedings. Figure \ref{fig:results1} shows these records in relationship to the CHI year they appear in.  
\hspace{-15pt}Exponential trends seem to indicate a proliferation of records matching our search-terms in both, the full and main proceedings from 2013 onward. The most recent three iterations of the CHI conference from 2015-2017 account with 33 retrieved records (Figure \ref{fig:results1}, blue bars) for more than a one-third of the 83 publications in the main proceedings. Similarly, the other CHI papers from 2013-2017 cover with 38 records more than half of the 54 papers we retrieved with our search-query.

In the following analysis in this preliminary study, we will present the results of the analysis of the 83 CHI main proceedings we have retrieved and therefore not consider the 54 CHI extended abstracts at this time.

\subsection{Frequency Analysis of the Sci-fi Referrals}
\begin{figure}[ht]
  \centering
  \includegraphics[scale=0.53]{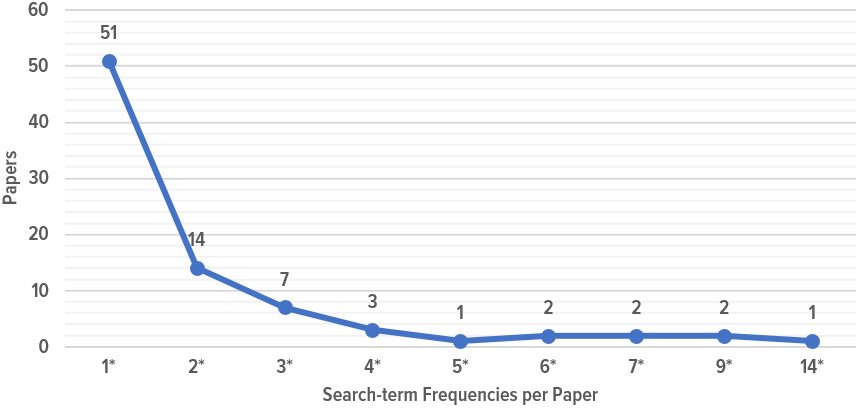}
  \caption{175 sci-fi referrals of our query in  83 papers}
  \label{fig:freq}
\end{figure}

Of the 175 sci-fi referrals we retrieve in the 83 papers in the main CHI proceedings (Figure \ref{fig:freq}), we find that 51 publications mention sci-fi as defined per our search query once (1*) within the full-text, a result consistent with related work, e.g. \cite{levin2014films}. 
\newline
The three publications with the highest frequencies of sci-fi referrals on a sentence-level analysis are: 
\begin{itemize}
    \item Blythe's \cite{Blythe:2014:RTD:2556288.2557098} 2014 CHI paper on imaginary abstracts with fourteen (14*) sci-fi referrals.
    \item A 1992 CHI panel discussion by Marcus, Norman, Rucker, Sterling and Vinge \cite{Marcus:1992} on sci-fi with nine (9*) sci-fi referrals.
    \item Tannenbaum's \cite{Tanenbaum:2012} 2012 CHI paper on steampunk with nine (9*) sci-fi referrals.
\end{itemize}

\hspace{-15pt}The earliest mention of sci-fi is found in Devaris' \cite{DeVaris:1982:IEH:800049.801795} 1982 paper on  information technologies and innovative workspaces, wherein he briefly compares (page 3):
\begin{quote}
    \textit{``[...] newly designed corporate office installations in Sweden, Switzerland, West Germany and Japan to the settings we would expect to see in a \textbf{science fiction} or space movie.''}
\end{quote}


\subsection{Coding Results of the Sci-fi in the 83 CHI Main Papers }
In summary, out of the 83 retrieved publications, 31 refer to sci-fi in trivial ways with regards to our research question.  For example, we retrieve papers which refer to sci-fi, but do in fact report on a usability evaluation of a video game set in a sci-fi universe.  Other publications we consider as false-positives mention a movie recommendation system, wherein sci-fi is a genre among others in addition to publications which mention sci-fi in the publication references, but do not discuss them in a meaningful way in the remainder of the paper. 

From the remaining 52 publications, 21 papers mention specific sci-fi in the context of the referral, such as movies (e.g. \textit{Terminator}, \textit{Star Wars}, \textit {Doctor Who}, authors (e.g. Isaac Asimov, H.G. Wells, Bruce Sterling), or sci-fi writings and novels (e.g. \textit{Dune}, \textit{Schismatrix}, \textit{Shaping Things)} and 31 papers refer unspecific sci-fi. Table \ref{tab:acmsearch2} presents the overview of the 83 CHI main papers we have coded as false-positives or which refer either, unspecific or specific sci-fi. 
\begin{table}
\centering
\caption{Retrieved papers, publication years and coding}
\label{tab:acmsearch2}
\begin{center}
\begin{tabular}{l|l|c||c|c|c}
\hline
\textbf{CHI Series} & \textbf{Range}  &   \textbf{Records} & \textbf{False-Positives} & \textbf{Unspecific sci-fi} & \textbf{Specific sci-fi}\\
\hline
CHI PRO & 1982-2017 & 83 & 31 & 31 & 21 \\
CHI OTH & 1996-2017 &  54 & Not Analyzed & Not Analyzed & Not Analyzed       \\
\hline
Total   & 1982-2017 & 137 & Not Analyzed & Not Analyzed & Not Analyzed \\    
\hline
\end{tabular}
\end{center}
\end{table}
\newline
In cases of multiple types of mentions, a specific sci-fi referral (movie, show or story/writing) was given priority over an unspecific sci-fi referral in our coding. For example, if a paper mentions a science fiction movie and then proceeds to introduce the movie \textit{Minority Report} in the text, we have coded that paper as referring specific sci-fi.

In order to exemplify our coding, we introduce in the following sections selected examples of our coding. As we have retrieved 175 sci-fi referrals of our search-terms across all 83 CHI main papers (see Figure \ref{fig:freq}, page \pageref{fig:freq}), we do provide overviews of the full coding in Table \ref{tab:types} (page \pageref{tab:types}) and Table \ref{details} (page \pageref{details}) in the latter part of the paper. In addition to the overviews, we introduce below selected, representative examples for each code accordingly.

\subsubsection{False-Positives:}We find 31 instances of CHI papers which we evaluate as false-positives with regards to overall importance of the sci-fi referral in relationship to the paper focus and our research question. For instance Kim, McNally, Norooz and  Druin's \cite{Kim:2017:ISR:3025453.3025572} study on adult internet searching habits refers our search-terms in quotes from study participants, for instance (page 4953 and page 4955):
\begin{quote}
    \textit{``I know several websites I would go to find \textbf{science fiction} books for kids;''} 
\newline
\newline
    \textit{``Like P17, several adult search users had domain knowledge related to their hobbies or leisure activities (e.g., travel, pop music, \textbf{science fiction} book).''} 
\end{quote}
In other examples of false-positives, sci-fi is referred in the context of emotions and digital games, for instance in Bopp, Mekler and Opwis's \cite{Bopp:2016:NEP:2858036.2858227} article describing the preferences of study participants (page 2998):
\begin{quote}
    \textit{``The three most popular genres were RPGs and action RPGs (n = 109), followed by adventure and action adventure games (n = 95), as well as strategy and real time strategy games (n = 71). Popular genres for other media were, for example, \textbf{science fiction}, comedy, drama, fantasy, and thrillers.''}
\end{quote}
Although sci-fi is mentioned within the full-text of the paper in these examples, we did assess these referrals as false-positives with respect to the research focus of the paper under review and the context of our content analysis. 
\subsubsection{Unspecific sci-fi:}
Krekhov, Emmerich, Babinski and Kr{\"u}ger \cite{Krekhov:2017:GPV:3025453.3025641} refer unspecifc sci-fi and prior studies as means to derive gestural interactions. Quoted below is an example of a sci-fi referral which did have a stronger connection to the research theme of the paper under scrutiny in our coding (page 5285):
\begin{quote}
   \textit{``Another interesting approach to generate comprehensible gestures is to rely on popular \textbf{science fiction} movies. Filmmakers have to create futuristic interactions that the audience should be able to understand intuitively.''}
\end{quote}
Yet serving as another example of an unspecific sci-fi referral, Bardzell, Bardzell and Stolterman \cite{Bardzell:2014:RCD:2556288.2557137} refer to sci-fi theory in their inquiry into critical design (page 1958): 
\begin{quote}
    \textit{``Following \textbf{science fiction} theorists in distinguishing between `cognitive speculation' and `mere fantasy,' in which the former is speculation grounded in a rigorous understanding of the past and present and used to project possible futures, and the latter is mere escapism, this radio seems to be more on the order of fantasy than serious speculation, and therefore the design \textit{qua} critical design is unsuccessful.''}
\end{quote}
\subsubsection{Specific sci-fi:} A paper by Reeves \cite{Reeves:2012:EUC:2207676.2208278} on the future of ubicomp is one example of a case where we identified specific sci-fi in our qualitative review (page 1577):
\begin{quote}
\textit{``Underkoffler would go on to produce a `Minority Report' gestural interface, which he demoed at Technology, Entertainment, Design (TED) 2010 conference. It is important to see this within the context of a wider historical trend, i.e., that of substantial feedback between the \textbf{science fiction} imagination and technological development.''}
\end{quote}
In this example, Reeves \cite{Reeves:2012:EUC:2207676.2208278} mentions specific sci-fi by naming the well-known sci-fi movie \textit{Minority Report} and links it to future technological advances.

\subsection{Research Themes where Sci-fi and HCI Research Intersect}
The 52 publications which we consider relevant were further analyzed and - based on the research theme and focus of the individual paper - placed into five, mutually exclusive thematic categories. In order to exemplify the thematic coding of the sci-fi in our analysis, we will introduce yet again selected examples of the coding for each thematic area. Then, we will provide summaries and details of the coding and analysis of the usage of sci-fi in the CHI main proceedings. 
\subsubsection{Coding Examples of Theoretical Design Research:}
For instance, Bauer and Kientz \cite{Bauer:2013:DSD:2470654.2466258} refer sci-fi in their CHI '13 paper on ideation through scenario-based design (page 1958):
\begin{quote}
    \textit{``Other, less traditional ideation methods include bodystorming, futures and alternative news, and \textbf{science fiction} prototyping. [...] \textbf{Science fiction} prototyping aims to use science fiction writing as a way of inspiring futuristic ideas. These methods often require the designers to generate the ideas and present them to users for feedback, and thus DesignLibs makes a unique contribution by enabling potential end users
to easily generate and contribute design ideas.''}
\end{quote}
Among others, Blythe \cite{Blythe:2014:RTD:2556288.2557098} extensively writes about design fiction and refers specific sci-fi in the context of theoretical design research, for example (page 706):
\begin{quote}
    \textit{``Literary techniques such as pastiche have been used to place concept designs in different fictional worlds, some from \textbf{science fiction} e.g. 1984 or A Clockwork Orange, but also other cultural contexts like Agatha Christie’s Miss Marple stories or the Simpsons.''}
\end{quote}
\subsubsection{Coding Examples of Human-Robot Interaction and AI:}
To begin with, Pereira, Prada and Paiva's \cite{Pereira:2014} 2014 case study on social presence in human-agent interactions leads with the observation that sci-fi robots are becoming reality (page 1449):
\begin{quote}
   \textit{``\textbf{Science-fiction} films or books have long included characters such as intelligent computers, robots and androids that evoke the same type of social responses from the audience or the reader. With the evolution of technology, these \textbf{science-fiction} visions are now becoming a reality, and new interactive techniques and devices are designed to evoke social responses from users.''}
\end{quote}
Bucci, Cang, Valair, Marino, Tseng, Jung, Rantala, Schneider and MacLean \cite{Bucci:2017:SCC:3025453.3025774} similarly write that (page 3683):
\begin{quote}
\textit{``Companion robots that once existed only in \textbf{science fiction} are quickly becoming part of our present reality.''}
\end{quote}
Takayama, Groom and Nass \cite{Takayama:2009} utilize a piece of dialogue from the seminal sci-fi movie \textit{2001: A Space Odyssey} as title in their 2009 CHI paper on human agents: \textit{``I'm Sorry, Dave: I'm Afraid I Won't Do That: Social Aspects of Human-Agent Conflict''}. The authors refer first to unspecific sci-fi, and then to the sci-fi movie \textit{2001: A Space Odyssey} as well as sci-fi author Issac Asimov (page 2099):
\begin{quote}
\textit{``On the other hand, Asimov’s Second Law of Robotics places obedience to humans above everything (including the robot’s self-preservation) other than harm to humans. This notion is echoed in \textbf{science fiction}: The disobedience of HAL, the computer from 2001, led to the death of his crew mates.''}
\end{quote}
\subsubsection{Coding Examples of New Interactions:}
As the first example of the theme of New Interactions, we identify Marcus et al. CHI '92 panel \cite{Marcus:1992}, quoted below, which had the clear aim to identify synergies between sci-fi authors and HCI researchers (page 435):
\begin{quote}
    \textit{``This plenary panel will explore ideas about future user  interfaces, their technology support, and their social  context as proposed in the work of leading authors of  \textbf{science fiction} characterized as the Cyberpunk  movement.''}
\end{quote}
Elmqvist, Henry, Riche and Fekete \cite{Elmqvist:2008:MSF:1357054.1357263} do acknowledge in 2008 that the source of inspiration for their paper on a novel interaction technique - space deformation - is a specific, 50-year old sci-fi story by sci-fi author Frank Herbert (page 1341):
\begin{quote}
    \textit{``The inspiration for the technique in this paper comes from Frank Herbert’s classic \textbf{science-fiction} novel Dune from 1965, where interstellar space travel is performed through a process known as `folding space'.''}
\end{quote}
As one last example, Grandhi, Joue and Mittelberg's \cite{Grandhi:2011:UNI:1978942.1979061} study on touchless gestural interfaces introduces the topic of novel interactions by means of a popular, specific sci-fi movie (page 821): 
\begin{quote}
    \textit{``However, unlike touch gestures, touchless gestures remain largely a notion developed in science fiction (as depicted in the popular \textbf{sci-fi} movie Minority Report) and have only been implemented to a limited degree in a few proof-of concept research applications [...].''}
\end{quote}
 
\subsubsection{Coding Examples of Visions of Computing and HCI:}In this thematic area, we code papers which refer sci-fi in conjunction with technology, agency, power and utopian and dystopian visions of the future.
For example, autonomous cars are subject to Lee, Joo and Nass's \cite{Lee:2014} study, which refers to the US sci-fi show \textit{Knight Rider} (page 3631):
\begin{quote}
    \textit{``Driverless car KITT in a famous American TV series is no longer \textbf{science fiction}.''}
\end{quote}
Mankoff, Rode and Faste's \cite{Mankoff:2013} paper on forecasting, HCI and futurism refers to a variety of sci-fi, for example (page 1629):
\begin{quote}
    \textit{``These authors ask us to dive into a world where \textbf{science fiction} bleeds into reality and the future suddenly seems surprisingly near and uncertain.''}
    \end{quote}
Later on, the authors \cite{Mankoff:2013} proceed to state a clear synergy of sci-fi, HCI and the technological future (pages 1632-1633): 

\begin{quote}
    \textit{``A critical examination of forecasts that arise from monitoring can inform HCI. An example is the analysis of \textbf{science fiction} to gain new insights into how ubiquitous computing technologies should engage with bureaucratic structures [...] by examining the cultural biases present in visions of the future (such as the powerful male who must escape from feminine control in movies such as The Truman Show), we create an opportunity to choose whether or not to reinforce them.''}\end{quote}
As a last example in this category, we present a 2017 case study \cite{Oh:2017:UVT:3025453.3025539} of 22 participants who attended a men-versus-machine Go match. Interviews with the attendants suggest a future view of society, technology and AI, which seems mainly shaped by a variety of famous sci-fi movies (page 2526):
\begin{quote}
    \textit{``When asked about their thoughts and impressions of the term AI, most of the participants described experiences of watching \textbf{science fiction} movies. They mentioned the specific examples, such as Skynet from Terminator (1984), Ultron from The Avengers (2015), Hal from 2001: A Space Odyssey (1968), sentient machines from The Matrix (1999), and the robotic boy from A.I. Artificial Intelligence (2001).''}
\end{quote}

\subsubsection{Examples of Human-Body Modification or Extension:}
\begin{figure}[ht]
  \centering
  \includegraphics[scale=0.51]{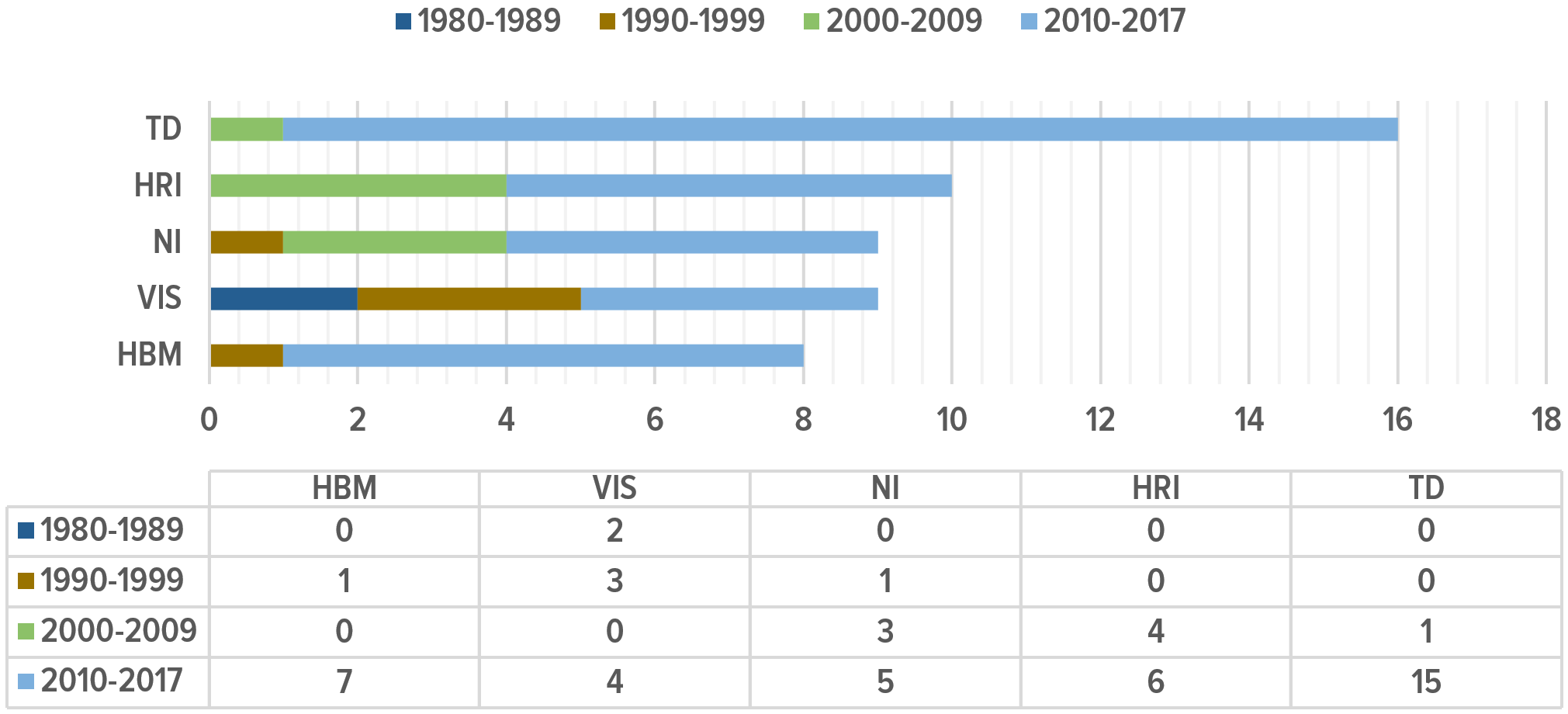}
  \caption{Research themes per decade (n=52) where sci-fi and HCI intersect in a meaningful way  in the CHI main proceedings.}
  \label{fig:acmsearch2}
\end{figure}

In the nascent research area of Human-Body Modification or Extension, we find referrals to sci-fi, for instance by Massimi, Odom, Banks and Kirk \cite{Massimi:2011:MLD:1978942.1979090} (page 993):
\begin{quote}
\textit{``The idea of using technology to prevent death entirely, or to `speak from beyond the grave' has been a motif in \textbf{science fiction} for decades, but has begun to actually occur in current systems.''}
\end{quote}
Jamison-Powell, Briggs, Lawson, Linehan, Windle and Gross \cite{Jamison-Powell:2016:PIL:2858036.2858504} observe a proliferation of posthumous messaging applications and mention unspecific sci-fi in their paper (page 2926): \begin{quote}
\textit{``Terming the second over-arching concept `Transcendence' may initially sound rather futuristic and, even, reminiscent of \textbf{science-fiction} when applied to the subject of death and technology. However, our analysis shows that these posthumous services do, indeed, transcend.''}
\end{quote} 
Permanent modification of the human body is subject to Heffernan, Vetere and Chang's \cite{Heffernan:2016:YPH:2858036.2858392} CHI '16 paper, who outline through specific sci-fi movies and shows the visual depictions of such interfaces in the past (all quotes, page 1799):
\begin{quote}
    \textit{``This section describes the use of insertables as seen in popculture and the recent leap from \textbf{science fiction} into reality.''}
\end{quote}
\begin{quote}
    \textit{``Insertable devices have accented \textbf{science fiction} for decades, from the Cyber Men of Doctor Who (1966) to The Terminator (1984). The 1970s TV series The Six Million Dollar Man and The Bionic Woman saw humans rebuilt; in their universes we had the technology.''}
\end{quote}
\begin{quote}
\textit{``While the above is not yet technically possible, the concept of insertables is no longer contained to the boundaries of \textbf{science fiction.}''}
\end{quote}
Britton and Semaan \cite{Britton:2017:MCV:3022198.3024939} draw similar analogies of sci-fi and the human-machine symbiosis (page 2499): 
\begin{quote}
    \textit{``From The Terminator to Fit Bit, the fascination with merging body and technology has played a persistent role in science and \textbf{science fiction}.''}
\end{quote}
In summary, we code the 52 CHI main papers into the five research themes as summarized below:

\begin{enumerate}
\item \textbf{Theoretical Design Research (n=16):} Out of the 52 relevant publications, 16 papers which refer sci-fi in publications on critical design research, ideation and design fiction or sci-fi prototyping as the examples presented in \cite{Blythe:2014:RTD:2556288.2557098}. 
\item \textbf{Human-Robot Interaction and AI (n=10):}  Out of the 52 relevant publications, 10 refer sci-fi in papers on human-robot or human-agent interaction and agency, artificial intelligence, natural language interfaces and ethics, for example \cite{Pereira:2014}.
\item \textbf{New Interactions (n=9):}  Out of the 52 relevant publications, 9 refer sci-fi in papers on novel interfaces and interaction modes (gestural, haptical, shape-changing, multi-modal) as the example presented in \cite{Grandhi:2011:UNI:1978942.1979061}.
\item \textbf{Visions of Computing and HCI (n=9):} Out of the 52 relevant publications, 9 refer sci-fi in publications on technology in conjunction with society, agency and power, for instance autonomous cars and systems or dystopian visions of ubiquitous computing in the conflict zone of privacy and security, e.g.
\cite{Mankoff:2013}.
\item \textbf{Human-body Modification or Extension (n=8):}  Out of the 52 relevant publications, 8  refer sci-fi in papers on DIY cyborgs or on-body fabrication of artifacts, implants, insertables and technologies for the digital afterlife, for example \cite{Heffernan:2016:YPH:2858036.2858392}.
\end{enumerate}
Table \ref{tab:types} summarizes the distribution of the 83 papers across our coding categories, while Table~\ref{details} on page~\pageref{details} shows the coding of each individual paper. 
\begin{table}
\centering
\caption{Coding overview of the 83 main CHI papers}
\label{tab:types}
\begin{center}
\begin{tabular}{l|l|c|l}
\hline
\textbf{Code} & \textbf{Coding} & \textbf{Range of Papers} & \textbf{Coding}  \\
\hline
FP & False-Positives & 1995-2017 & 31  \\
TD & Theoretical Design Research & 2000-2017 & 16 \\
HRI & Human-Robot Interaction and AI & 1996-2017 & 10 \\
NI & New Interactions & 1992-2017 & 9 \\
VIS & Visions of Computing and HCI & 1982-2017 & 9  \\
HBM & Human-Body Modification & 1999-2017  & 8 \\
\hline
 & \textbf{Total}   & \textbf{1982-2017} &  \textbf{83} \\
\hline
\end{tabular}
\end{center}
\end{table}
\newline Excluding the 31 false-positives for illustration purposes, Figure \ref{fig:acmsearch2} shows the 52 main CHI proceedings we coded in one of the final five research themes. In order to indicate trends, the 52 papers have been further grouped into the decade the papers were published in.

\begin{table}[h]
\centering
\caption{Coding details of the 83 main CHI papers}
\label{details}
\begin{tabular}{|c|c|c|c|c|c|c|c|}
\hline
 & \textbf{FP} & \textbf{TD} & \textbf{HRI} & \textbf{NI} & \textbf{VIS} & \textbf{HBM} & \textbf{TOTAL} \\ \hline
1982 &  &  &  &  & 2 &  & 2 \\ \hline
1992 &  &  &  & 1 & 1 &  & 2 \\ \hline
1995 & 3 &  &  &  &  &  & 3 \\ \hline
1996 & 2 &  &  &  &  &  & 2 \\ \hline
1997 &  &  &  &  & 2 &  & 2 \\ \hline
1999 & 1 &  &  &  &  & 1 & 2 \\ \hline
2000 &  & 1 &  &  &  &  & 1 \\ \hline
2001 & 1 &  &  & 1 &  &  & 2 \\ \hline
2002 & 1 &  &  &  &  &  & 1 \\ \hline
2003 & 1 &  &  &  &  &  & 1 \\ \hline
2004 & 1 &  &  &  &  &  & 1 \\ \hline
2006 & 1 &  &  &  &  &  & 1 \\ \hline
2008 & 1 &  & 2 & 1 &  &  & 4 \\ \hline
2009 & 2 &  & 2 & 1 &  &  & 5 \\ \hline
2010 & 1 &  &  &  & 1 &  & 2 \\ \hline
2011 &  &  &  & 1 &  & 1 & 2 \\ \hline
2012 &  & 2 &  &  &  &  & 2 \\ \hline
2013 & 3 & 2 &  &  & 1 & 1 & 7 \\ \hline
2014 & 2 & 2 & 2 & 1 & 1 &  & 8 \\ \hline
2015 & 3 & 1 &  & 2 &  &  & 6 \\ \hline
2016 & 4 & 2 & 3 &  &  & 4 & 13 \\ \hline
2017 & 4 & 6 & 1 & 1 & 1 & 1 & 14 \\ \hline
\textbf{TOTAL} & \textbf{31} & \textbf{16} & \textbf{10} & \textbf{9} & \textbf{9} & \textbf{8} & \textbf{83} \\ \hline
\end{tabular}
\end{table}

\section{Findings and Conclusions}
In order to address the lack of a study into the usage of sci-fi referrals in HCI research, our research question examined if, and how, general and specific sci-fi is used the ACM Conference on Human Factors in Computing Systems. In this preliminary study, we identified 137 records in the CHI proceedings which refer sci-fi - as defined with our six-term search and retrieval query. We also  qualitatively analysed the 83 records indexed in the main CHI proceedings which fitted out inclusion criteria.

In our qualitative review, we further identified 52 publications which discuss sci-fi in a meaningful way with regards to our research question and in addition, five broader research themes which intersect with sci-fi in papers from 1982-2017.
Specifically, we explored the usage of sci-fi, the specific type of sci-fi referral, the search-term frequency distribution of sci-fi referrals, and the associated technologies of sci-fi in scientific publications. 

In general, it appears that sci-fi becomes more acknowledged in the CHI conference in recent years. The majority of relevant publications (37 out of 52) in our analysis are published in the 2010s onwards indicating an upsurge of sci-fi papers in recent years.

Our research also provided a better understanding of the evolutionary uses of sci-fi in HCI research in the case of the CHI main track. With regards to the research themes in focus on the papers we analysed, we observed that \textit{Visions of Computing and HCI} of the future in conjunction with sci-fi are referred from the 1980s onwards. In contrast, \textit{Human-Robot Interaction and AI} publications have been referring sci-fi in the 2000s forward through for instance,  movies and fictional robots to explore robot ethics or the public perception of AI. 

Since the 2010s, there are emerging research areas in computing research, predominantly the thematic area of \textit{Theoretical Design Research}, but as well the areas of \textit{New Interactions} or \textit{Human-Body Modification}. In these thematic areas, sci-fi movies, shows or stories do provide an inspiration for the foremost and upcoming HCI challenges of our time, for example through the discussion of shape-changing interfaces, implantables or digital afterlife ethics.

This research was of an exploratory nature. By taking the CHI proceedings as a pilot dataset for analysis, we afforded ourselves the possibility to apply our tentative methodology. In the process, we learned its strengths and limitations. All in all, our methodology proved useful as we were able to identify specific sci-fi through a contextual review of the retrieved papers. We aim to use it to conduct further more fully fledged studies where we will investigate how sci-fi is used in HCI.

\section{Discussion and Future Work}

We acknowledge that our open, qualitative coding of the uses of sci-fi within the respective publications might be subjective as it might lack external validity and overall reliability. Although we discussed both, our coding scheme  and the five general research themes thorough, in some cases the mutual exclusive coding of the sci-fi in the paper under review was challenging.

For instance,  \cite{Oh:2017:UVT:3025453.3025539} shares aspects of both, the theme of Human-Robot Interaction and AI and Visions of Computing and HCI. In another, similar example, we find a paper \cite{Lee:2014} which refers specific sci-fi, namely KITT from \textit{Knight Rider}. KITT is arguably an AI of some sorts embedded in a vehicle, hence the paper presents results of a study on human driver's loss of agency in the emergence  of `partially intelligent' cars in the near future. 

To establish a basic reliability of our method, we did first find common ground through open coding 25\% of the sample independently. Then we consolidated and, after one author finished the review of the remaining papers, conducted a final discussion of the coding between the raters. Still, we did not calculate a Krippendorff's alpha, respectively Fleiss' kappa, in this study, a limitation we aim to address in future studies in order to validate our coding scheme. 

Moreover, the full-text query of the dataset we used did lead us to specific sci-fi authors, movie titles, and short stories in the context of the retrieved paper. Although, that is a reasonable expectation, we might have missed relevant publications not matching our initial six-term search query. Direct full-text searches for movie titles \cite{levin2014films}, such as \textit{``Terminator''} or \textit{``Avatar''}, can generate hundreds of false-positives which complicate the qualitative coding process.

This trade-off between recall and precision, or full-text versus metadata search and retrieval, is a classic dilemma in information retrieval \cite{Beall2008,Hemminger2007,Lin2009}. In this study, we decided to search full-text for 'sci-fi' and its synonyms and therefore might not have caught all potential referrals of the concept. While this is a trade-off we will reconsider in a future instance of this ongoing research, we believe our results are interesting and shed light on remarkable  trends in HCI research.

Our analysis shows that starting in the 2010s, there is an upsurge of retrieved publications, specifically from 2013 in the CHI extended abstracts, and from 2016 onwards in the main CHI proceedings. On the other hand, these initial results can only be of provisional nature and should be normalized against the distribution of all papers per year in the main CHI proceedings. Such a comparison might allow for a better insight into an above-average occurrence of sci-fi related papers. This is then an aspect of this preliminary study that warrants further investigation.
 
As of yet, our analysis  does not include the other 54 CHI records, among those the records from the CHI extended abstracts  (e.g. other work-in-progress papers, alt.chi contributions, plenary sessions, workshops, etc.). We aim to qualitatively review these 54 other records in the future, and plan to explore in more detail the specifics of the sci-fi referred in HCI research.
 
Beyond the extension of this study, future analysis will apply similar and more specific searches in a wider range of literature, conferences and venues. Exploring sci-fi referrals in conferences other than CHI, for example HCII, UIST, INTERACT, SIGGRAPH or even the IEEE \textit{Xplore} Digital Library might allow for comparative assessments of both, research themes and cultural bias across conferences and periodicals in order to corroborate or refute the results of this study. 
 
In the future, we aim to collect a larger, yet filtered, sample that is generated from cross-cultural and preferably, univocal, sci-fi titles and keywords in order to replicate and validate this pilot study. We also aim to investigate in more details the uses of specific types of sci-fi, for instance particular sci-fi movies, novels or characters and aim to investigate as well the uniformity - or diversity - of the sci-fi material consulted, refereed and discussed in HCI research. 
 
Based on our analysis and findings, a last important future goal of this research effort will be to extrapolate any lessons that allow researchers and practitioners to understand why sci-fi is used and how the sci-fi supports the arguments of the papers where it is cited, to ultimately provide guidance on future research in this area. 

We speculate that the explicit referral of sci-fi in HCI research represents a fraction of the actual inspiration and impact it has had on HCI research. Our future studies will therefore aim to conduct qualitative interviews with HCI researchers in order to investigate that claim and assess the reasons and reservations of authors in computer science, who refer, or rather decide consciously to not refer to sci-fi in their research output.

\bibliographystyle{splncs03}
\bibliography{sigproc}

\end{document}